\documentclass[prl,preprintnumbers,twocolumn,showpacs]{revtex4}

\usepackage{amsfonts,amsmath,mathrsfs}

\def \a {\alpha}
\def \b {\beta}

\def \d {\delta}

\def \f {\phi}

\def \et {\eta}

\def \ka {\kappa}
\def \la {\lambda}

\def \z {\zeta}
\def \m{\mu}
\def \n{\nu}

\newcommand{\printsize}{
      \headheight=10pt                              
     \topmargin=-1cm \headsep=0.5cm
      \oddsidemargin=-0.9cm \evensidemargin=-0.9cm  
      \textheight=24truecm \textwidth=18truecm      %
      \setlength{\columnsep}{20pt}                  
}

\printsize

\def \be{\begin{equation}}
\def \ee{\end{equation}}
\def \bse{\begin{subequations}}
\def \ese{\end{subequations}}
\def \bd{\begin{displaymath}}
\def \ed{\end{displaymath}}
\def \ba{\begin{eqnarray}}
\def \ea{\end{eqnarray}}
\def \bas{\begin{eqnarray*}}
\def \eas{\end{eqnarray*}}

\def \pt{\partial}
\def \lag{\mathscr{L}}
\def \nb{\nabla}
\def \dd{\mathrm{d}}
\def \DD{\mathcal{D}}
\def \ex{\mathrm{e}}
\def \mm{\int\!\dd^4 x}

\newcommand \diag[1]{\mathrm{diag}[{#1}]}
\newcommand \lie[1]{\pounds_{#1}}
\newcommand \tr[1]{\mathrm{Tr}\left\{{#1}\right\}}

\newcommand \ord[1]{\mathcal{O}({#1})}
\newcommand \cbr[1]{\left({#1}\right)}
\newcommand \sbr[1]{\left[{#1}\right]}
\newcommand \pbr[1]{\left\{{#1}\right\}}

\usepackage{txfonts}

\begin{document}

\title{Gauge Dependence of Gravitational Correction to Running of Gauge Couplings}

\author{Artur R. Pietrykowski}
\email{Hearthie@ift.uni.wroc.pl}
\affiliation{Institut of Theoretical Physics, University of Wroclaw, 50-204 Wroclaw, Plac Maxa Borna 9, Poland}


\begin{abstract}
Recently an interesting idea has been put forward by Robinson and Wilczek
that incorporation of quantized gravity in the framework
of Abelian and non-Abelian gauge theories
results in a correction to the running of gauge coupling and, as a consequence,
increase of the grand unification scale and asymptotic freedom.
In this Letter it is shown by explicit calculations that this
correction depends on the choice of gauge.
\end{abstract}

\maketitle

Since all the fundamental interactions in the standard model of particle physics are
perturbatively renormalizable, this property became a criterion
for any theory which is to be called a fundamental one.
It is, however, a well known fact that the quantized General Relativity is perturbatively
nonrenormalizable theory \cite{DeWittDuffChrist, tHooftVeltman},
and as such cannot aspire to being the fundamental
theory of gravitational interactions. Incorporation of other
fields: scalar \cite{tHooftVeltman}, spinor \cite{Deser1}
and Yang-Mills \cite{Deser2, Deser3} does not improve this result.
It is interesting in this context to inquire whether the quantized gravitational field
affects the gauge coupling of the Yang-Mills fields. From refs. \cite{Deser2, Deser3} it
follows immediately that the one loop $\b$-function remains unaltered.
Nevertheless it has been suggested \cite{OdintsovPercacci} that
in analogy to the Gross-Neveu model interacting with the 2-dimensional Liouville theory
one should expect a "gravitational dressing"
of the gauge coupling in $S\!U\!(N)$ gauge theory
coupled to the conformal sector of the 4-dimensional quantum gravity.
In fact, the "gravitationally dressed" gauge coupling
is of the form $g^2(E)=g^2/(1+b_0 g^2 \a^{-1} \log E^2),$
where $\a$ is the anomalous scaling dimension and $b_0$ is the usual factor
depending on the matter content of the gauge group. It has been found that
for the Einstein gravity in the Standard Model $\a = 1.13$ \cite{OdintsovPercacci}. This makes
the unification scale increase by a factor of the order of 2.

There is another point of view that stems from the analogy
to the description of the soft pion scattering by means of
an effective field theory. Namely, we can treat the general relativity as
an effective field theory that results from
as yet unknown theory of quantum gravity.
This approach casts new light on the problem of whether quantized gravity
alters the behavior of gauge couplings. Thus the results of refs. \cite{Deser2, Deser3}
should be reinterpreted in this context.
Robinson and Wilczek \cite{Robinson:Wilczek:2005}
have found that there is a quantum gravitational correction to running of gauge coupling
in the case of the $S\!U(N)$ theory minimally coupled to gravity.
This correction has the form of an additive extra term to a Callan-Symanzik
$\b$ function, namely
$$
\b(g,E)=-b_0 g^3/(4\pi)^2 - 3 (E/M_P)^2 g/\pi.
$$
The negative
sign of this correction works in the direction of asymptotic freedom when the gauge coupling
is evaluated at higher energies. The additive form of this correction signifies that it is
valid for all gauge couplings. This stems form the fact that gravitons carry no gauge charge.
In the case of $U(1)$ theory the running of gauge coupling is
$\a(E)=\a(0)\exp(-3\tfrac{\ka^2}{(4\pi)^2}E^2).$ The conclusions of \cite{Robinson:Wilczek:2005}
have been recently considered in the context of extra-dimensional models \cite{Gogoladze:Leung},
where the gravitational scale is lower then $M_P.$ Estimates made in \cite{Gogoladze:Leung}
allow these gravitational corrections to be tested in the LHC experiments.

The result of \cite{Robinson:Wilczek:2005} has been derived in the framework of the background field method.
The background metric is taken to be the Minkowski metric.
It means that the Einstein equations of motion cannot be satisfied. In this case, as many authors
have pointed out \cite{DeWittDuffChrist,Grisaru:Nieuwenhuizen:1975,GaugeDependence,tHooft:1975},
if the background field method is applied to a system with a gauge symmetry
which is off the mass shell then, in general, one should expect counterterms
to depend on the choice of gauge (although Kallosh has shown \cite{Kallosh:1974}
that in the case of pure Yang-Mills theory counterterms are gauge independent
even if a system is off-shell). In particular in Ref. \cite{Grisaru:Nieuwenhuizen:1975}
it has been shown that in the case of quantized gravity interacting with a scalar field,
counterterms depend on the choice of gauge parameter if the system is off the mass shell,
whereas on-shell this dependence disappears. Therefore it should be examined
whether the result obtained in \cite{Robinson:Wilczek:2005} for the
gravitational correction to running of gauge coupling is gauge independent.
This is the main goal of this Letter.



\textit{Form of the correction.} ---
Since we are concerned with the problem of gauge dependence of gravitational corrections it is worth
following the main steps that led to the result of Ref. \cite{Robinson:Wilczek:2005}.
For the sake of simplicity and transparency of arguments to be used we will consider
Abelian case, \textit{i.e.} the Maxwell field.
In order to quantize the Einstein-Maxwell system we use the technique of the background field
described in detail in \cite{tHooft:1975}. Let us thus consider
the action of the Einstein-Maxwell theory which is of the
following form
\ba
\label{emaction}
\nonumber
S &=& S_E+S_M
\\
&=&-\mm\sqrt{-\bar g}
\Big(\frac{1}{\ka^2}R(\bar
g)+\frac{1}{4}\bar{g}^{\m\a}\bar{g}^{\n\b}\bar{F}_{\m\n}\bar{F}_{\a\b}\Big),
\ea
where $\ka=\sqrt{16\pi G}$ (the minus sign in front of the integral
indicates the convention we have used for Ricci tensor).
Fields $\bar g_{\m\n}, \bar A_{\m}$ may be written as sums
of background and quantum fields, respectively,
\be
\label{defpol}
\bar g_{\m\n}=g_{\m\n}+\ka h_{\m\n},\quad \bar A_\a = \ka^{-1}A_\a+a_\a .
\ee
For the sake of convenience we have redefined the Maxwell field so as $A_\a \to \ka^{-1}A_\a$,
but the final result will be quoted with the original definition restored.
We develop the action up to the second order in quantum fields.
We obtain terms quadratic in quantum fields.
Now, it is necessary to fix the gauge.
The transformation laws which leave the action (\ref{emaction}) invariant are
\bse
\ba
\label{transh}
\d_\xi h_{\m\n}&=&2\nb_{(\m} \xi_{\n)}+ \ka\ \lie{\xi}h_{\m\n},
\\
\label{transa} \d_\xi a_\a &=& \lie{\xi}A_\a + \ka\ \lie{\xi}a_\a,
\ea
\ese
where $\lie{\xi}$ is Lie derivative. They correspond to infinitesimal coordinate
transformations ${x'}^\m=x^\m+\ka \xi^\m(x).$
The Abelian gauge transformation for Maxwell field reads
\bd
\d_\la a_\m = \nb_\m \la, \quad
\d_\la h_{\m\n}=0.
\ed
So far, we have done all the computations in curved spacetime.
Further considerations, however, will be performed under the assumption,
that the background spacetime is flat. An immediate consequence of this is
that the Einstein equations are not satisfied, and as a consequence,
the energy-momentum tensor does not vanish.
However, we assume the Maxwell equations to be satisfied.
In order to reproduce the result of Ref. \cite{Robinson:Wilczek:2005}
we take the following $\z$-parameter dependent class of gauge conditions
for the graviton field, called $R_\z$-gauge
and the usual Lorentz gauge for the photon field
\be
\label{rgauge}
C_\m(h) - \z\  F_\m^{\ \a} a_\a = 0 ,\quad \pt_\a a^\a = 0,
\ee
where
\be
\label{harmonic}
C_\m(h) \equiv \pt^\n h_{\n\m}-\tfrac{1}{2}\pt_\m h , \quad
h \equiv \et^{\a\b}h_{\a\b}.
\ee
Later on we shall specify the gauge taking $\z = 1$. Having chosen the gauge we may proceed according to
the Faddeev-Popov method which amounts to an addition to the Lagrangian of quantum fields the gauge-breaking terms
as well as the vector and the scalar ghost Lagrangians.

Having chosen the gauge the Lagrangian of quantum fields may be rearranged so that it
can be treated as a function of multi-component real scalar field
$\f^T = (h_{\m\n}, a_\a)^T,$ namely
\be
\label{operator}
\lag_{\f}=\tfrac{1}{2}\f^T(W\pt^2+\bar N^\m \pt_\m + \bar M) \f \equiv \tfrac{1}{2}\f^T\mathcal{A}\f,
\ee
where, from the point of view of the renormalization procedure the only interesting
matrices are $\bar N^\m$ and $\bar M$. $\bar N^\m$
is antisymmetric and contains graviton-photon cross terms linear in $F$,
\textit{i.e.} the background field Maxwell strength tensor.
$\bar M$ is symmetric and consists of the off-diagonal elements of the form $\pt F$
and elements $\ord{F^2}$ in both graviton and photon sectors.
It is worth noting that an appearance of the terms $\ord{F^2}$ in the photon sector is due to
the specific gauge (\ref{rgauge}).

Computation of the one-loop quantum correction to the coupling constant
amounts to the evaluation of the functional integral in the formula for the effective action defined by
\be
\nonumber S_{\mathrm{eff}}[A]
=-i\log\pbr{\int\DD \f\DD \la^{\ast}\DD \la\ \ex^{i S[\et,A,\f,\la,\la^*]}},
\ee
where $\la$ is the Maxwell ghost field. As a result of computation one obtains that the only insertion
that corrects the classical Maxwell action comes from the logarithm of functional determinant,
which is the trace of logarithm of the operator in Eq. (\ref{operator}). It amounts to
\be
\label{rw:tra}
-\frac{1}{2}\tr{\log(-\mathcal{A})} \approx -3i\frac{1}{(4\pi)^2}(\Lambda^2-E^2) S_M+\ord{F^4}.
\ee
As this determinant is ultraviolet divergent we subtract the divergence at a reference energy $E_0$.
The effective action then reads
\bd
S_{\mathrm{eff}} \approx
\sbr{\frac{1}{e^2}-\frac{\ka^2}{e^2}\frac{3}{(4\pi)^2}(E_0^2-E^2)}S_M.
\ed
We have retrieved the original field dimension,
and we have extracted explicitly the coupling constant \textit{i.e.} $A_\m \to (\ka/e) A_\m$.
For values $E$ infinitesimally close to $M$ one may extract from the above action the one-loop $\b$ function
\bd
\b(e,E)=-3\frac{\ka^2}{(4\pi)^2}e E^2.
\ed
Hence, a formula for running coupling constant
is $\a(E)=\a(E_0)\exp\big[-3\tfrac{\ka^2}{(4\pi)^2}(E^2-E_0^2)\big].$
This formula constitutes a basis for the conclusion drawn in \cite{Robinson:Wilczek:2005} that
for the energy scale $E \gg M$ coupling constant vanishes,
and hence QED is an asymptotically free theory when a quantized
gravitational interaction is taken into account.
It is interesting, however, whether this result would have
changed if we had taken a different gauge condition.



\textit{Verification of gauge dependence.} ---
In order to verify the gauge dependence we take instead of $R_\z$
a class of gauges characterized by a parameter $\xi$.
For this class $\xi=1$ corresponds to the harmonic gauge (both classes are equivalent for $\z =0$ in Eq. (\ref{rgauge})
and $\xi = 1$). From now on we call this class the generalized harmonic gauge.
After an addition of the gauge-braking term
\be
\label{gblag}
\lag_{\rm GB} = \frac{1}{2\xi}C_\m^2(h),
\ee
[cf. Eq. (\ref{harmonic})] the Lagrangian for quantum fields
may be rearranged in a similar manner as before. Now, its multi-component form is the following
\be
\label{lagfi}
\lag_\f = \tfrac{1}{2}\f^T[D(\xi)\pt^2+N^\m\pt_\m+M]\f
= \tfrac{1}{2}\f^T\mathcal{A}(\xi)\f.
\ee
where
\be
\label{si}
D(\xi) = W + \cbr{1-\frac{1}{\xi}}\ \diag{K(\pt),0}.
\ee
The first matrix in Eq. (\ref{si}) appeared already in Eq. (\ref{operator}). The second one has the
diagonal element in graviton part only. It is a combination of differential operators
that come from the form of the gauge-breaking Lagrangian in Eq. (\ref{gblag}). The matrices
$N^\m$ and $M$ are antisymmetric and symmetric respectively as their counterparts in Eq. (\ref{operator}).
However their matrix elements differ from those
in Eq. (\ref{operator}) as a result of the choice of gauge. Namely, the off-diagonal elements in $N^\m$
are now combinations of terms linear in $F$. The off-diagonal elements of the matrix $M$
are the same as in in Eq. (\ref{operator}). The difference is in the diagonal element of the photon sector
which is zero. This difference is crucial for the one-loop computations.

As before, the computation of the one-loop correction is reduced to an evaluation of the logarithm
of functional determinant of the operator given in Eq. (\ref{lagfi}).
Since we are concerned with terms $\ord{F^2}$ that correct the coupling constant
of the Maxwell action we expand this logarithm of the determinant up to terms of this order.
The only insertion that we expect comes from the matrix $M$, and the square of
the matrix $N^\m$. Hence, one gets
\begin{widetext}
\be
\label{deta}
\tr{\log[-\mathcal{A}(\xi)]}\approx -\tr{D^{-1}(\xi)\frac{1}{-\pt^2}M}
-\frac{1}{2}\tr{D^{-1}(\xi)\frac{1}{-\pt^2}N^\m\pt_\m D^{-1}(\xi)\frac{1}{-\pt^2}N^\n\pt_\n},
\ee
where $D^{-1}(\xi)$ is the inverse of the operator in Eq. (\ref{si}).
Evaluation of the first term in Eq. (\ref{deta}) yields
\be
\label{trm}
-\tr{D^{-1}(\xi)\frac{1}{-\pt^2}M}\approx i\sbr{6 +
3(1-\xi)}\frac{\Lambda^2-E^2}{(4\pi)^2} S_M.
\ee

Computation of the second term in Eq. (\ref{deta}) by virtue of the
definition of the operator in Eq. (\ref{si}) boils down to evaluation of
three traces, namely
\bse
\ba
\label{tr1}
\tr{D^{-1}(\xi)\frac{1}{-\pt^2}N^\m\pt_\m D^{-1}(\xi)\frac{1}{-\pt^2}N^\n\pt_\n}
&=&\tr{\frac{1}{-\pt^2}W^{-1} N^\m\pt_\m\frac{1}{-\pt^2}W^{-1} N^\n\pt_\n}
\\
\label{tr2}
& & - 2\cbr{1-\frac{1}{\xi}}\tr{\frac{1}{-\pt^2}W^{-1}
N^\m\pt_\m\frac{1}{-\pt^2}K(\pt)N^\n\pt_\n}
\\
\label{tr3}
& & + \cbr{1-\frac{1}{\xi}}^2
\tr{\frac{1}{-\pt^2}K(\pt)N^\m\pt_\m\frac{1}{-\pt^2}K(\pt)N^\n\pt_\n}.
\ea
\ese
Inspection of these traces reveals that only two of them, i.e., Eqs. (\ref{tr1}) and (\ref{tr2})
give a contribution, as the third one is equal to zero. Thus, the result of the computations yields
\be
\label{tr123}
-\frac{1}{2}\tr{D^{-1}(\xi)\frac{1}{-\pt^2}N^\m\pt_\m D^{-1}(\xi)\frac{1}{-\pt^2}N^\n\pt_\n}
\approx -i\sbr{6 + 3(1-\xi)}\frac{\Lambda^2-E^2}{(4\pi)^2} S_M.
\ee
\end{widetext}
Notice, that this result has the same form as in Eq. (\ref{trm}) but with the opposite sign.
This means that both of the insertions to Eq. (\ref{deta}) compensate one another.
Therefore as the final result we obtain
\be
\label{traxi}
\tr{\log[-\mathcal{A}(\xi)]}\approx 0 + \ord{F^4}.
\ee
We conclude that the one-loop effective action does not depend on the gauge parameter $\xi$
even when the system is off-shell. We also see that as the trace in Eq. (\ref{traxi}) vanishes,
there is no one-loop correction to running of gauge coupling. Therefore this result
differs from that obtained in Eq. (\ref{rw:tra})
in $R_\z$ gauge for $\z=1$ [cf. Eq. (\ref{rgauge})].
Note also that lack of any one-loop correction is in agreement with power counting arguments
given in \cite{Deser2}.



\textit{Conclusions.} --- In this paper we have investigated whether the
gravitational correction to the $\b$ function
which has been calculated in \cite{Robinson:Wilczek:2005} depends on the choice of gauge.
Therefore we have first reproduced the result of \cite{Robinson:Wilczek:2005}
in the simplest case, \textit{i.e.} in the Maxwell theory. In order to attain this we have
taken $R_\z$ type of gauge (\textit{cf.} Eq. (\ref{rgauge})). We have then
repeated all the calculations taking the generalized harmonic gauge defined
in Eq. (\ref{gblag}) and parameterized by parametr $\xi$.
We have performed all the calculations without specifying the value of $\xi$,
in order to check whether the final result depends on this parameter. We have found
that this is not the case. However, the result we have obtained reveals lack of
gravitational correction up to one-loop order. It is consistent with
power counting arguments of Ref. \cite{Deser2}. This, however,
means that the conclusions drawn in \cite{Robinson:Wilczek:2005}, and pertaining
to a nontrivial correction in $R_\z$ gauge to the $\b$ function are specific to this gauge,
and are devoid of any physical meaning.
As a final remark, let us notice that if we applied the dimensional regularization
we would obtain no gravitational correction, since in the case of a massless fields this
method employs the logarithmic divergence and neglects the polynomial one.

The author would like to thank Professor Z. Haba for inspiration and
many useful discussions, and Professor S. Odintsov for interesting
conversation and valuable remarks.

{\it Note added.} --- During the review process, we were forwarded
the following response from Robinson and Wilczek: "Pietrykowski
has calculated the one-loop graviton corrections to the effective
action of quantum Einstein-Maxwell theory and found a zero result,
in contrast to our original non-zero result. He calculated in four
dimensions with harmonic gauge, using integration by parts to
appropriately symmetrize between the gravitational and gauge
sectors of the action. We have now redone our calculation in a
general gauge and arbitrary dimension using Pietrykowski's
symmetrization procedure. This gauge is the $R_\xi$-like gauge
from our paper with arbitrary value of $\xi$. (We originally
choose $\xi=1$. Pietrykowski chooses $\xi=0$, although he denotes
this as $\zeta=0$ since he uses $\xi$ for a different gauge
parameter.) For $\xi=0$ in four dimensions, we agree with
Pietrykowski's result of zero correction to the effective action,
although zero correction is not generic. The important point is
that this check revealed (to us) unexpected gauge dependence. We
thought we had a proof that this did not occur, and do not agree
that arguments based on dimensional regularization imply {\it a
priori} that the corrections must vanish; obviously, we will have
to track down a flaw, either in our logic or in our algebra."

\end{document}